\documentclass{JINST}

\title{Investigation of gamma ray detection performance of thin LFS scintillator with MAPD readout}

\author{E.Guliyev$^a$, F.Ahmadov$^b$,G.Ahmadov$^b$,A. Sadigov$^b$, Z.Sadygov$^b$,N. Anfimov$^b$, Z. Krumshtein$^b$, A. Olshevski$^b$,V. Shvetsov$^b$,V. Zhezher$^b$,R. Madatov$^c$\\
\llap{$^a$}LLR-Ecole Polytechnique,CNRS/In2p3,\\
   91128 Palaiseau, France CEDEX\\
\llap{$^b$}Joint Institute for Nuclear Research,\\
  141980 Dubna, Russia\\
\llap{$^c$}Institute of Radiation Problems,\\
  AZ-1143 Baku, Azerbaijan\\
  E-mail: \email{elmaddin@llr.in2p3.fr}}

\abstract{Prototype of gamma ray detector based on Micro Pixel Avalanche Photodiodes (MAPD) with high pixel density 15000 pixel/$mm^2$, optically coupled to Lutetium Fine Silicate (LFS) scintillator has been developed. The detection performance investigated in the range of energy 59.6 - 662 keV at room temperature. The results of measurements are presented in this work.}


\begin{document}

\section{Introduction}

Gamma ray detectors plays an important role to detection and spectroscopy of the emitted photons of unstable nucleus, nuclear reactions and high energy physics experiment. The performance of gamma ray detectors puts a series of requirements on detector active material and readout electronics. For instance, the main requirements of detector material to have a high light yield and fast decay time, radiation hard and high photon detection efficiency, small size, high sensitivity for single photons in wide spectral energy range, insensitivity to magnetic field, low noise, dynamic range for detector readout electronics together with photo-sensor . The scintillation crystal Lutetium Fine Silicate (LFS-8) fits very well those requirements in small size with the combination of MicroPixel Avalanche Photodiode (MAPD). The performance of gamma ray detector based on LFS-8 scintillator with MAPD photo-sensor investigated in the energy range from 59.6 keV to 662 keV. MAPD and scintillator LFS-8 manufactured by Zecotek Company were used ~\cite{zekotek}.

\section{Experimental setup and results}

In this study, LFS-8 scintillating crystal with dimensions 3 x 3 x 0.5 $mm^3$ produced by Zecotek Company, was tested. The surface of crystal polished and was shielded with 5 $\mu $m thin aliminium film ~\cite{workshop}. The light luminescence yield compared to NaI(Tl) is very high, about 80\% (main properties of LFS-8 crystal are presented in Table 1).  

\begin{table}[h]
\centering
\caption{Main properties of LFS-8.}
\begin{tabular}{|l|c|}
\hline
Parameter & LFS \\
\hline
Density & 7.13 g/$cm^3$\\
\hline
Light yield & 30000 ph/MeV\\
\hline
Decay time & 12-25 ns\\
\hline
Attenuation length & 2.6 cm\\
\hline
\end{tabular}
\label{tab:template}
\end{table}

\begin{figure}[h]
\setlength{\abovecaptionskip}{-60pt}
 \centering
  \includegraphics[width=0.70\textwidth,clip]{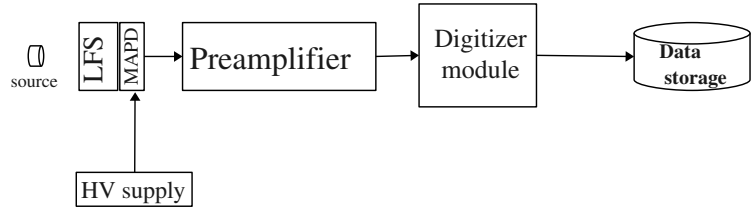} 
  \caption{ The scheme of the experimental setup to study the performance of the gamma ray detector base on thin LFS scintillator with MAPD readout ~\cite{workshop}.}
  \label{ex_setup}
\end{figure}

The LFS-8 scintillation crystals assembly with the MAPD to measure the energy response of radioactive sources. These photo-sensors have several advantages: insensitivity of magnetic field and a high pixel density, 15000 pixels/$mm^2$ ~\cite{zair}, high gain $5.5\times10^4$ and photon detection efficiency 25-30\% at room temperature in a wide wavelength range at operation voltage ~\cite{mapd}. The number of pixels in per $mm^2$ is almost 10 time higher and less dark count compare to Hamamatsu silicon based MPPC (Multi Pixel Photon Counter)  ~\cite{hamam}. Dark current is around of 70 nA at operation voltage 94 V. Capacitance of MAPD reaches up to 120 pF ~\cite{mapd1}.
The block-diagram of experimental setup is depicted in Figure ~\ref{ex_setup} and contains photo-detector, charge sensitive preamplifier, the digitizer module.  The surface of LFS-8 scintillator was placed at a distance of 1 cm from the gamma ray source. The preamplifier output signals were directly digitized by the 12-bit resolution and 250 MS/s sampling rate of CAEN DT5720B digitizer module (~\cite{caen}) and the digitized traces were stored on disk for further analysis. 
 
\begin{figure}[H]
 \centering
  \includegraphics[width=0.60\textwidth,clip]{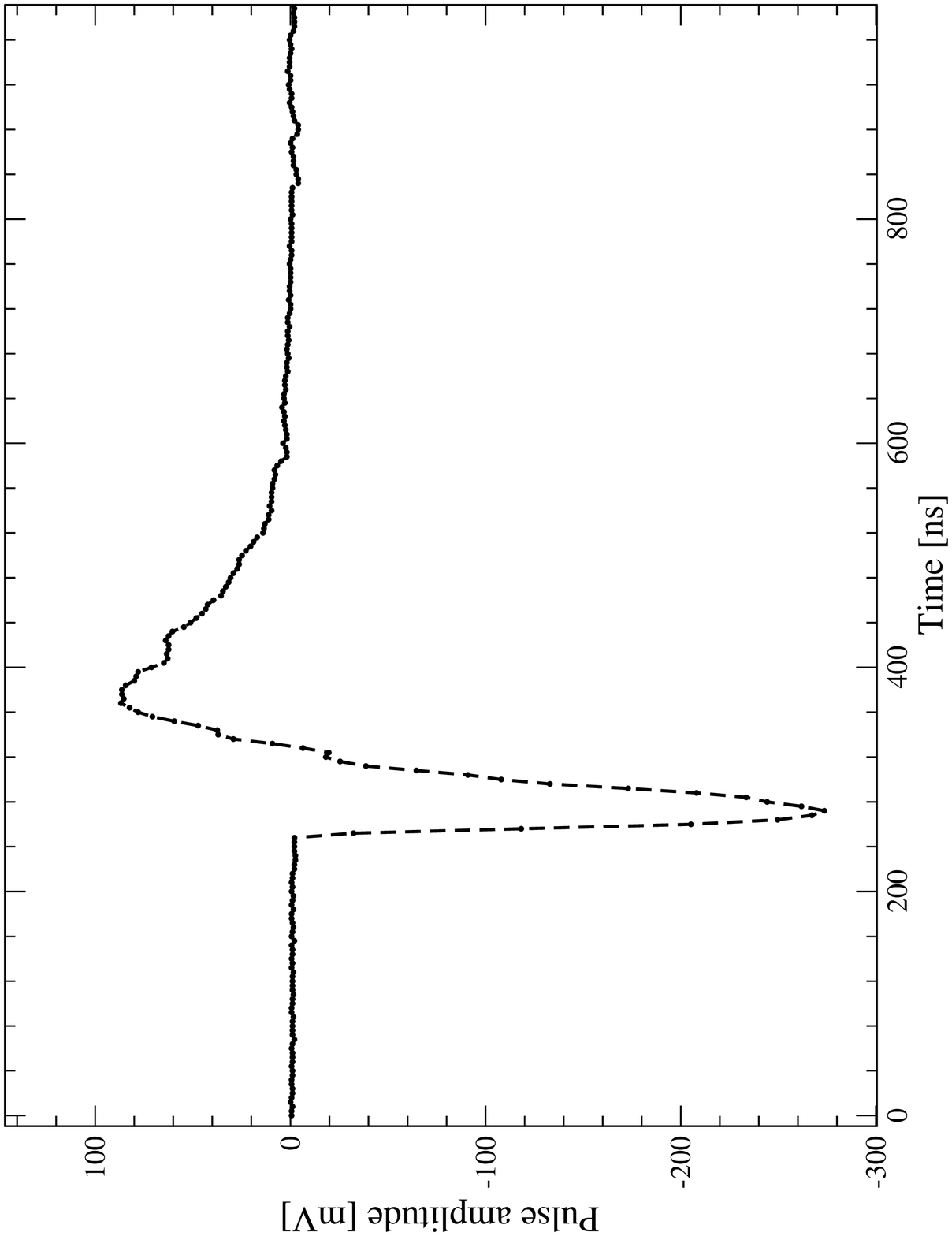} 
  \caption{ The recorded trace from the preamplifier output.}
  \label{ex_trace}
\end{figure}

The data acquisition (DAQ) was triggered when the gamma ray of the point source hits the detector. For the events, we took the amplitude of the obtained pulse to obtain about the energy. The baseline is estimated from the left side of preamplifier waveform (Figure ~\ref{ex_trace}). 

\begin{figure}[H]
 \centering
  \includegraphics[width=0.60\textwidth,clip]{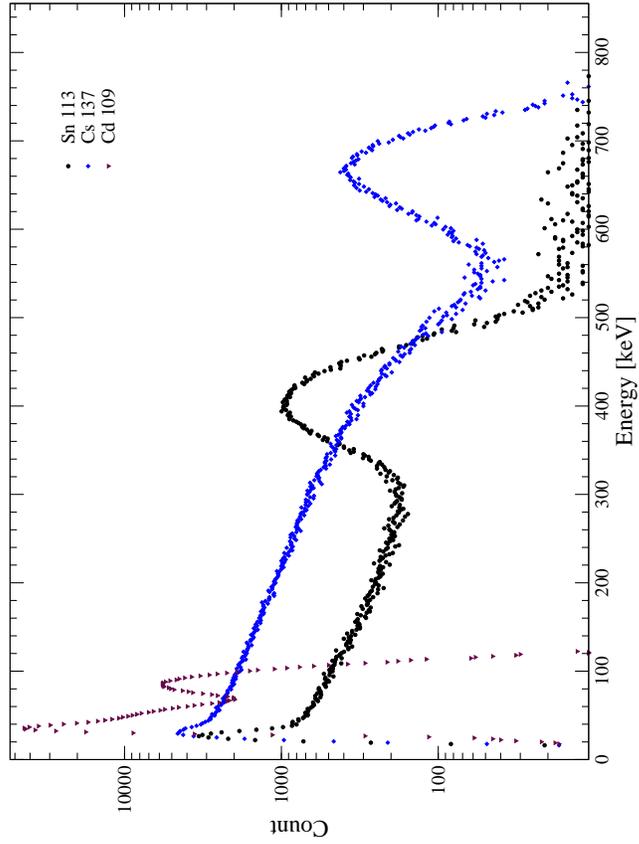} 
  \caption{ The energy spectrum for different gamma ray source in the LFS-8 scintillator coupled to MAPD readout.}
  \label{spectrum}
\end{figure}

	The details of used sources is shown in Table 2. The corresponding spectrum of that sources presented in Figure ~\ref{spectrum}. As shown in Figure ~\ref{spectrum}, the combination of LFS-8 with MAPD gives possibility to detect the low rate gamma rays (Sn-113 70Bq ).  
 
\begin{table}[h]
\caption{Activity and gamma ray energy of radioactive sources.}
\centering
\begin{tabular}{|l|c|c|}
\hline
Radionuclide & Activity, kBq & Energy, keV \\
\hline
Am-241 & 101 & 59.6\\
\hline
Cd-109 & 241   & 88\\
\hline
Co-57 & 4.7 & 122\\
\hline
Ce-139 & 0.22 & 165.9\\
\hline
Sn-113 & 0.07 & 391.7\\
\hline
Cs-137 & 104 & 662\\
\hline

\end{tabular}

\label{tab:template}
\end{table}

\begin{figure}[H]
 \centering
  \includegraphics[width=0.60\textwidth,clip]{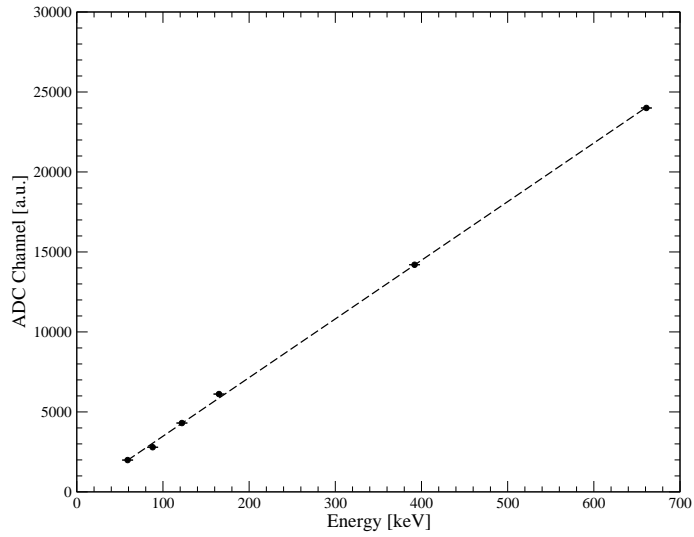} 
  \caption{ The calibration curve between measured ADC channels and known gamma ray energy. The uncertanities for measured peak positions to smaller.}
  \label{linearity}
\end{figure}

The calibration curve between measured ADC channels and known gamma ray energy was given in  Figure ~\ref{linearity}. The uncertainties for measured peak positions are very small.

\begin{figure}[H]
 \centering
  \includegraphics[width=0.60\textwidth,clip]{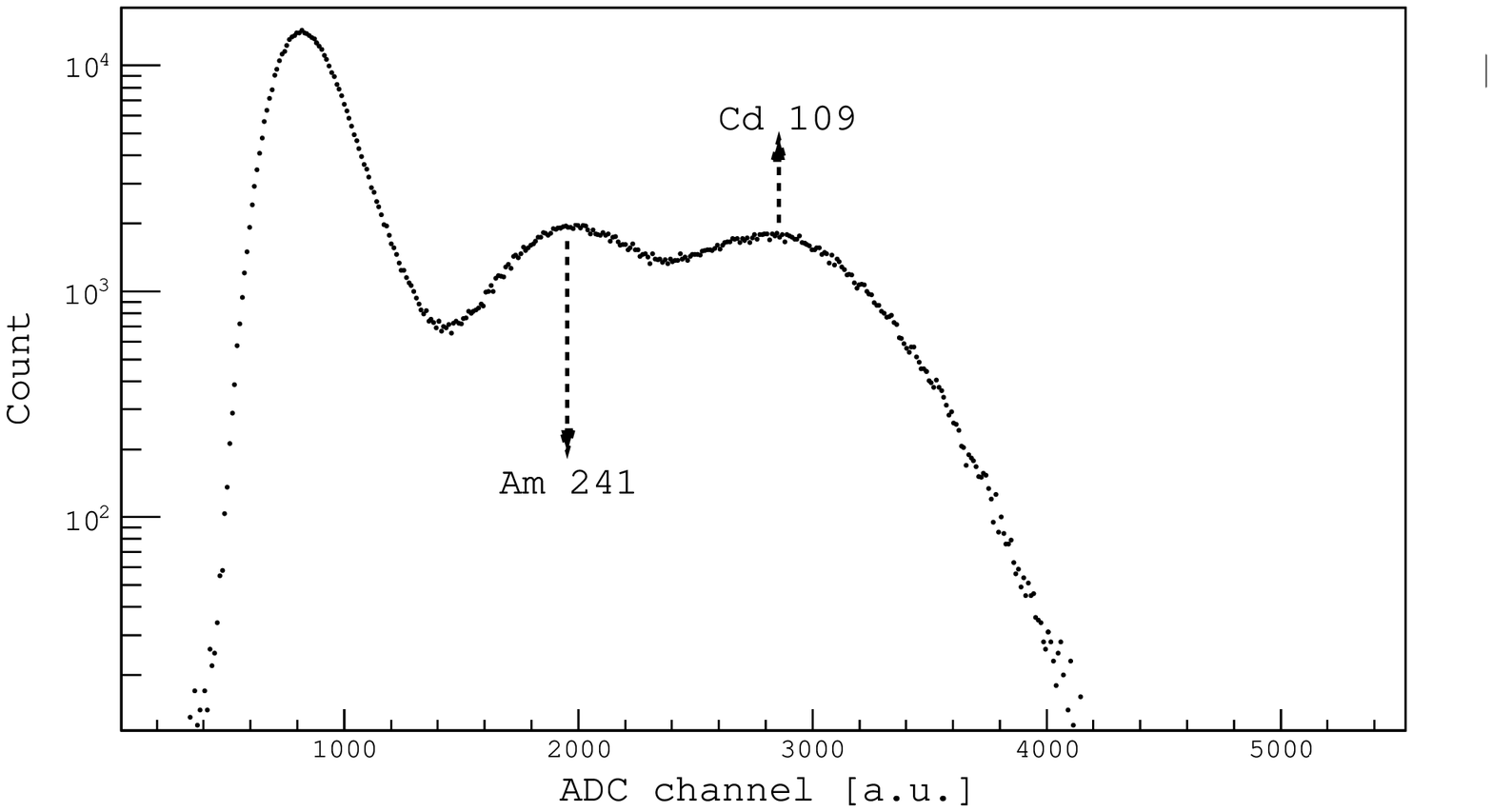} 
  \caption{ The energy spectr measured at same time with Am-241 and Cd-109 gamma ray sources. Its clear seems that 3 x 3 x 0.5 $mm^3$ size of scintillator with 3 x 3 $mm^2$ MAPD can distinguish low energy 59.6 keV and 88 keV gammas.}
  \label{mixum}
\end{figure}

	There is one measurement we confirmed to see that, the separation of the gamma rays, where energies to closed each other. For this measurement Am-241 and Cd-109 sources placed together in front of the LFS-8 scintillator. From the Figure ~\ref{mixum} we can see separated gamma rays from different sources, that measured at same time.

\section{Conclusion}

The series of study was shown that LFS-8 scintillator with MAPD photo-sensor readout are suitable for compact gamma ray spectroscopy. Due to high efficiency of MAPD photo-sensor and enough luminesence light yield of LFS-8 scintillator gives possibility to utilize such combination low-level radioactivity measurement in environmental radiation pollution.

\end{document}